\documentclass[conference]{IEEEtran}
\IEEEoverridecommandlockouts
\usepackage{amssymb}
\usepackage{graphicx}
\usepackage{epstopdf}
\usepackage{gensymb}
\usepackage{color}
\usepackage{amsmath}
\usepackage{bm}
\usepackage{etoolbox}
\usepackage{cite}
\usepackage{array}
\usepackage{booktabs}
\usepackage{multirow}
\usepackage{comment}
\usepackage{subcaption}
\usepackage[ruled,vlined]{algorithm2e}
\usepackage{algorithmicx}
\usepackage{mathrsfs}
\usepackage{algpseudocode}
\usepackage{hyperref}
\hypersetup{
    colorlinks=true,
    linkcolor=blue,
    filecolor=magenta,      
    urlcolor=cyan,
}
 
\alglanguage{pseudocode}

\algnewcommand{\Initialize}[1]{%
  \State \textbf{Initialize:}
  \Statex \hspace*{\algorithmicindent}\parbox[t]{.8\linewidth}{\raggedright #1}
}

\usepackage{threeparttable}
\usepackage{mathtools}
\usepackage{hyperref}
\usepackage{pifont}
\usepackage{listings}

\usepackage{url}
\urldef{\mailsa}\path|{alfred.hofmann, ursula.barth, ingrid.haas, frank.holzwarth,|
	\urldef{\mailsb}\path|anna.kramer, leonie.kunz, christine.reiss, nicole.sator,|
	\urldef{\mailsc}\path|erika.siebert-cole, peter.strasser, lncs}@springer.com|    

\usepackage[utf8]{inputenc}
\usepackage[english]{babel}

\usepackage{amsthm}

\usepackage{soul}

\theoremstyle{definition}

\definecolor{R}{RGB}{0,0,150}

\theoremstyle{remark}

\ifCLASSINFOpdf

\begin{document}

\title{End-to-End Evaluation of Federated Learning and Split Learning for Internet of Things}


\author{
   \IEEEauthorblockN{Yansong Gao\IEEEauthorrefmark{1}\IEEEauthorrefmark{2}, Minki Kim\IEEEauthorrefmark{2}\IEEEauthorrefmark{3}, Sharif Abuadbba\IEEEauthorrefmark{1}\IEEEauthorrefmark{2}, Yeonjae Kim\IEEEauthorrefmark{2}\IEEEauthorrefmark{3}, Chandra Thapa\IEEEauthorrefmark{2}, \\ Kyuyeon Kim\IEEEauthorrefmark{2}\IEEEauthorrefmark{3}, Seyit A. Camtepe\IEEEauthorrefmark{2}, Hyoungshick Kim\IEEEauthorrefmark{2}\IEEEauthorrefmark{3}, and Surya Nepal\IEEEauthorrefmark{1}\IEEEauthorrefmark{2}}
  \IEEEauthorblockA{\textit{\IEEEauthorrefmark{1} Cyber Security Cooperative Research Centre}, Australia. \{garrison.gao; sharif.abuadbba; surya.nepal\}@data61.csiro.au}
   \IEEEauthorblockA{\textit{\IEEEauthorrefmark{2} Data61, CSIRO}, Syndey, Australia. \{minki.kim;chandra.thapa;seyit.camtepe;hyoung.kim\}@data61.csiro.au.}
   \IEEEauthorblockA{\textit{\IEEEauthorrefmark{3} Sungkyunkwan University}, Suwon, Republic of Korea.}
}


\maketitle

\begin{abstract}
Federated learning (FL) and split neural networks (SplitNN) are state-of-art distributed machine learning techniques to enable machine learning without directly accessing raw data on clients or end devices. In theory, such distributed machine learning techniques have great potential in distributed applications, in which data are typically generated and collected at the client-side while the collected data should be processed by the application deployed at the server-side. However, there is still a significant gap in evaluating the performance of those techniques concerning their practicality in the Internet of Things (IoT)-enabled distributed systems constituted by resource-constrained devices.

This work is the first attempt to provide empirical comparisons of FL and SplitNN in real-world IoT settings in terms of learning performance and device implementation overhead. We consider a variety of datasets, different model architectures, multiple clients, and various performance metrics. For the learning performance (i.e., model accuracy and convergence time), we empirically evaluate both FL and SplitNN under different types of data distributions such as imbalanced and non-independent and identically distributed (non-IID) data. We show that the learning performance of SplitNN is better than FL under an imbalanced data distribution but worse than FL under an extreme non-IID data distribution. For implementation overhead, we mount both FL and SplitNN on Raspberry Pi devices and comprehensively evaluate their overhead, including training time, communication overhead, power consumption, and memory usage. Our key observations are that under the IoT scenario where the communication traffic is the primary concern, FL appears to perform better over SplitNN because FL has a significantly lower communication overhead compared with SplitNN. However, our experimental results also demonstrate that neither FL or SplitNN can be applied to a heavy model, e.g., with several million parameters, on resource-constrained IoT devices because its training cost would be too expensive for such devices. Source code is released and available: \url{https://github.com/Minki-Kim95/Federated-Learning-and-Split-Learning-with-raspberry-pi}.
\end{abstract}

\begin{IEEEkeywords}
split learning, federated learning, distributed machine learning, IoT
\end{IEEEkeywords}


\section{Introduction}\label{sec:intro}

In recent years, the rapid proliferation of deep learning has led to stunning transformations in a wide variety of applications, including computer vision, disease diagnosis, financial fraud detection, malware detection, access control, surveillance, and so on~\cite{lecun2015deep,wang2017adversary,tang2016deep}. In general, deep learning models learn high-level invariant features by training them on rich data and using those features to solve the problems.



However, data can often be highly private or sensitive; for example, data collected from medical sensors~\cite{huang2019patient} and microphones~\cite{i2smicrophone} would be such cases. Consequently, users may resist sharing their data with service/cloud providers who are trying to build a deep learning model. On the other hand, the centralized data could be mishandled or incorrectly managed by service providers---e.g., incidentally accessed by unauthorized parties~\cite{amazonstranger}, or used for unsolicited analytics, or compromised through network and system security vulnerability---resulting in the data breach~\cite{shastri2019seven,biometricbreach}. Therefore, there is a demand for training a deep learning model without aggregating and accessing sensitive raw data resided in the client-side.

In this context, distributed learning techniques are being developed to tackle the above issues by training a joint model without accessing decentralized raw data held by clients in a distributed manner. Such techniques offer great potential for distributed system applications to reap the benefits from rich data generated/collected by IoT devices in distributed Internet of Things (IoT) architectures. Distributed learning techniques keep the data locally and utilize private data (e.g., medical records, voice records, and text inputs) during the learning process to reduce privacy leakage risks.     



In this paper, we consider two distributed learning techniques, namely federated learning (FL) and split neural network (SplitNN) (also referred to as split learning). FL is a well-known distributed learning technique~\cite{fedlearningMcMahan17,fed2}. In FL, a joint model is built through aggregating (e.g., averaging) models trained on each client's local data. SplitNN is a recently introduced distributed learning technique~\cite{gupta2018distributed,nopeeksplitNN}. In general, a neural network is split into two parts {\it vertically}. The first {\it few} layers belong to the client (e.g., IoT device), and the remaining layers belong to the server (e.g., cloud). The client and the server collaboratively train the whole network. There have been theoretical and empirical evaluations~\cite{zhao2018federated,yoon2020federated} on the FL. To the best of knowledge, however, there is no research analyzing the performance of SplitNN under diverse distributed data conditions. A recent study~\cite{singh2019detailed} compared both models in terms of communication efficiency only. However, they do not consider learning performance, such as model accuracy and convergence speed, especially when the data are imbalanced or non-IID (non-independent and identically distributed). We note that such imbalanced data situations would frequently happen in practice---e.g., IoT devices generate different types or/and sizes of data.

Furthermore, there is no empirical study on the end-to-end evaluation of FL and SplitNN for real-world IoT settings in terms of their implementation overhead, such as communication cost, power consumption, and training time. Indeed, as highlighted in~\cite{chen2019deep}, there is a demand to understand the deep learning performances on resource-constrained IoT/edge device hardware like  Raspberry Pi~\cite{raspberryPi}. Experimental results with real-world IoT devices would be useful for service providers considering the deployment of FL or SplitNN. Thereby, this paper aims to take the first step of empirically evaluate FL and SplitNN in real-world IoT applications. Main contributions/results of this work are summarized as follows:

\begin{enumerate}
    \item We are the first to evaluate SplitNN learning performance in terms of model accuracy and convergence under non-IID and imbalanced data distributions, and then compare it with FL under the same settings. Our \textit{empirical} results---up to simulated 100 clients---demonstrate that SplitNN exhibits better learning performance than FL under imbalanced data, but worse than FL under (extreme) non-IID data, indicating that SplitNN accuracy is sensitive to the characteristics of the distributed data.  
    \item We evaluate the applicability of mounting FL and SplitNN on resource-constrained IoT devices such as Raspberry Pi. Our intensive evaluation results suggest that complicated models (e.g., MobileNet) with several million parameters would be infeasible for such devices. For distributed {\it learning or training} with resource-constrained IoT devices, we recommend using a 1D CNN model with fewer parameters to deal with sequential data, which we thus have focused and extensively evaluated. An experimental video demo is available from \url{https://www.youtube.com/watch?v=x5mD1_EA2ps}.
    
    \item We take the first step toward fairly {\it training} performance comparisons between FL and SplitNN by mounting both on Raspberry Pi. We provide detailed performance overhead evaluations of training time, amount of memory used, amount of power consumed, communication overhead, peak power, and temperature to serve as a reference for practitioners. Under IoT scenarios in which the communication cost reduction is more important than the training time and energy consumption, FL seems to be a better option due to its significantly lower communication overhead compared with SplitNN.
    \item We are the first to empirically extend SplitNN for ensemble learning by exploiting the sequential learning process of SplitNN and the high computational power of cloud services to gain multiple models during the learning process of SplitNN while reducing the training expense.

\end{enumerate}

The remainder of this paper is organized as follows: Section~\ref{sec:background} details background on distributed learning models followed by our experiments and datasets used. We comprehensively evaluate the learning performance of both FL and SplitNN under both imbalanced and non-IID data distributions in Section~\ref{sec:learningEvaluation}. Section~\ref{sec:implementationEvaluation} mounts both FL and SplitNN on Raspberry Pi to empirically evaluate and compare their implementation overhead. We discuss the insight gained and provide future work in Section~\ref{sec:discussion}, followed by the conclusion in Section~\ref{sec:Conclusion}.

\section{Distributed Learning and Datasets}\label{sec:background}

We firstly describe background on FL, SplitNN, and ensemble learning techniques. Then we describe the datasets used in this work.

\subsection{Federated Learning}
\begin{figure}[!ht]
\begin{subfigure}{.5\textwidth}
  \centering
   \includegraphics[trim={0 2cm 0 0},clip,width=1\linewidth]{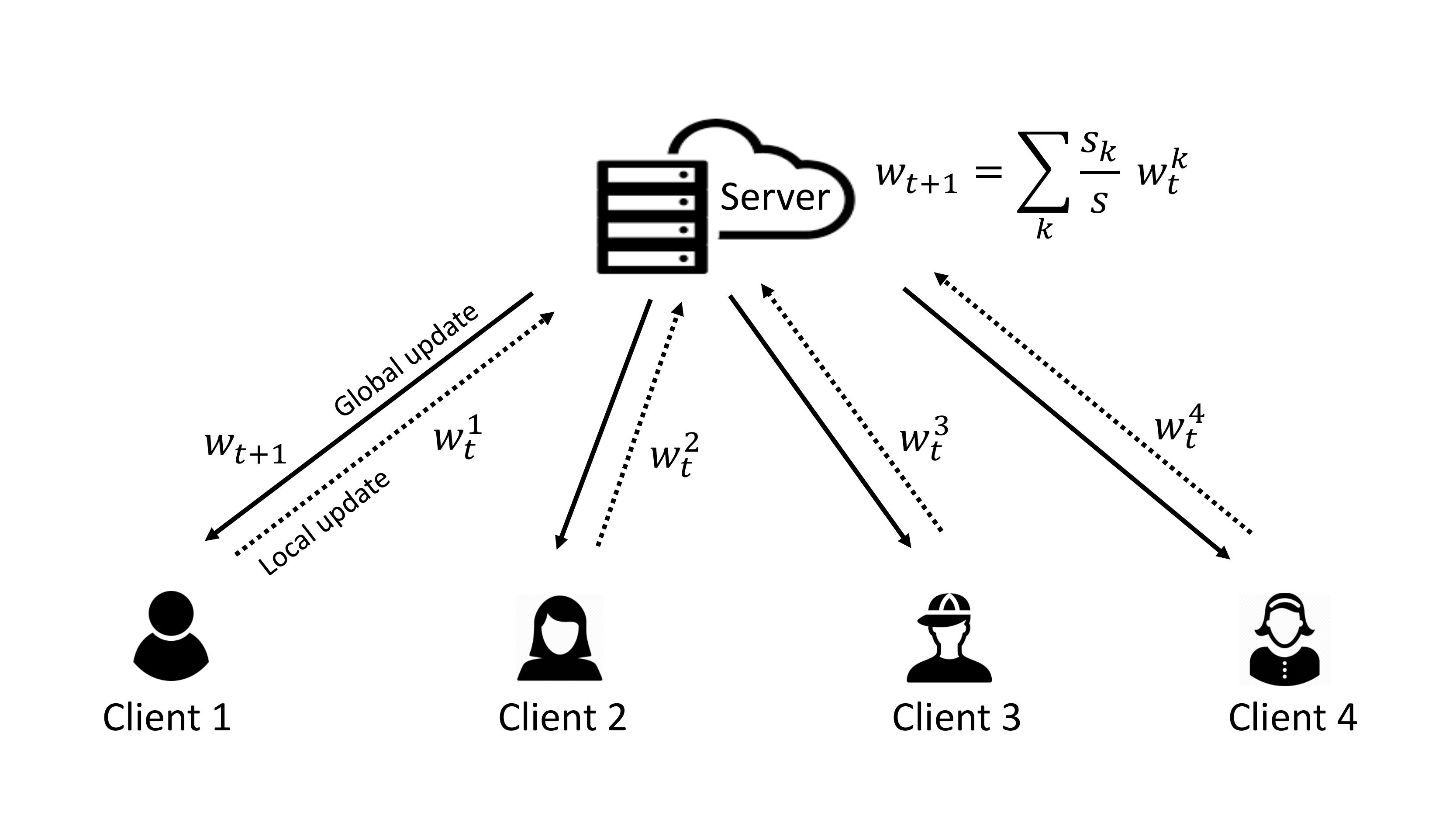}  
  \caption{FL with four clients}
  \label{fig:1}
\end{subfigure}
\begin{subfigure}{.5\textwidth}
  \centering
  \includegraphics[trim={0 2cm 0 0},clip=true,width=1\linewidth]{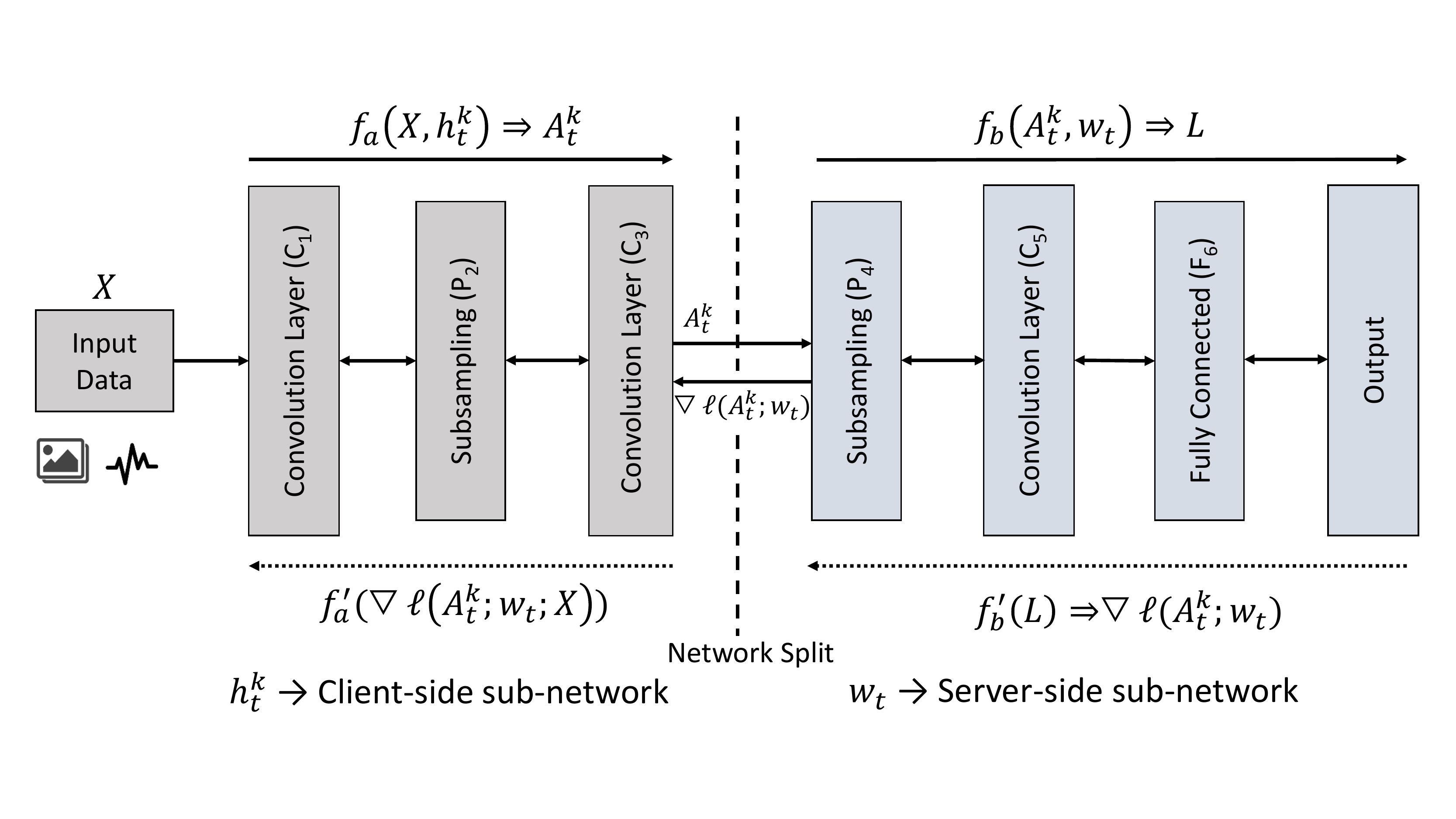}  
  \caption{SplitNN with seven layers}
  	\label{fig:2}
\end{subfigure}
\caption{Illustrated examples of FL and SplitNN.}
\label{fig:fig}
\end{figure}

The FL is illustrated in Fig.~\ref{fig:1}. During the training process, the server first initializes the global model $w_t$ and sends it to all participating clients. After receiving the model $w_t$, each client $k$ trains the global model on its local data---$s_k$ is the number of training samples held by client $k$ while $s$ is the total number of training samples across all clients. Afterward, each client returns the updated model $w_t^k$ to the server. The server then aggregates all those models to update the global model to get $w_{t + 1}$. The above process (often called \emph{round}) repeatedly continues until the model converges. 

According to~\cite{mcmahan2016communication}, instead of training the model with local data only one epoch, each client trains the local model for several epochs before sending it to the server in one communication round, which is referred to as \texttt{FedAvg} that is one commonly used method for FL optimization. Although \texttt{FedAvg} usually works well, specifically for non-convex problems, there are no convergence guarantees. FL may diverge in practical settings, especially if data are non-IID and imbalanced distributed across clients~\cite{li2019federated}.

\subsection{Split Learning}
Unlike FL, in which each client trains the whole neural network, The SplitNN~\cite{gupta2018distributed,nopeeksplitNN} divides a neural network model into at least two sub-networks, and then trains the sub-networks, separately, on distributed parties (e.g., client and server). The SplitNN is illustrated in Fig~\ref{fig:2}, where $C_3$ is the cut layer that divides the whole network into two sub-networks. The first sub-network $h_t$ is trained and accessed by the client; the second sub-network $w_t$ is trained and accessed by the server. Therefore, the server has no access to clients' sub-networks and data, which provides privacy protection. Besides privacy benefit, each client only needs to train a sub-network consisting of a few layers while most layers reside in the server. Therefore, as the other benefit, the client's computation load can be reduced.

The learning performance (e.g., model accuracy and convergence) of SplitNN has not been investigated yet when the data is non-IID or distributed in an imbalanced manner, which will be evaluated in this work. 

\subsection{Ensemble Learning}\label{sec:ensembleSetup}

Deep learning models are nonlinear methods that learn via a stochastic training algorithm, which will result in models suffering high variance. One can opt for multiple models for the same problem to address this, and their predictions are combined to make the final decision. This approach is called model averaging and belongs to a family of techniques, namely, ensemble learning. 

For SplitNN, the server interacts with each client a round-robin fashion~\footnote{To the best of our knowledge, there is no work on performing SplitNN among clients in a parallel manner upon the writing of this work.}. On the one hand, at a time, there is only one active client while the rest are waiting to be called. On the other hand, the server is computationally powerful with a cluster of GPUs. In this context, we can utilize the resourceful server to obtain multiple models by taking advantage of idle clients during SplitNN training. To be precise, we exemplify an ensemble learning process with two clients and two model architectures M$_1$ and M$_2$. The server first trains M$_1$ with client$_1$ on D$_1$ (data held by client$_1$) and trains M$_2$ with client$_2$ on D$_2$ (data held by client$_2$), which are performed simultaneously---both clients train at same time. Afterward, the server continues to train M$_1$ with client$_2$ on D$_2$ and trains M$_2$ with client$_1$ on D$_1$ in parallel. Once D$_1$ and D$_2$ are trained for both M$_1$ and M$_2$, one round is completed---equal to one global epoch for all the data across clients. In this manner, the server can train and consequentially gain two models compatible with the SplitNN training process. This assembling training can reduce computational overhead (e.g., time) to gain multiple models compared to train each model sequentially.

\subsection{Datasets}

Sequential data or time-series data is pervasively collected and processed by IoT devices. For example, people can order and purchase goods using speech commands at their voice assistant. Wearable medical sensors are used to monitor users' health status in real-time. Consequentially, we choose two such popular datasets: speech command (SC) and ECG for experimental evaluations, as summarized in Table~\ref{tab:Setup}. The SC is a personalized dataset, and the ECG is a medical dataset. Both datasets would be privacy-sensitive, where users are unwilling to share.  



\subsubsection{Speech Commands (SC)} This task is for speech command recognition. The SC contains many one-second .wav audio files: each sample has a single spoken English word~\cite{SClink}. These words are from a small set of commands and are spoken by a variety of different speakers. In our experiments, we use 10 classes: `zero', `one', `two', `three', `four', `five', `six', `seven', `eight', and `nine.' There are 20,827 samples where 11,360 samples are used for training, and the remaining samples are used for testing.


\subsubsection{Electrocardiogram (ECG)} 
MIT-BIH arrhythmia \cite{moody2001impact} is a popular dataset for ECG signal classification or arrhythmia diagnosis detection models. Following~\cite{kiranyaz2015real, li2017classification}, we collect 26,490 samples in total which represent 5 heartbeat types as classification targets: $N$ (normal beat), $L$ (left bundle branch block), $R$ (right bundle branch block), $A$ (atrial premature contraction), and $V$ (ventricular premature contraction). Half of them are randomly chosen for training, while the rest samples are for testing.


\begin{table}
	\centering 
	\caption{Datasets and Models.}
			\resizebox{0.5\textwidth}{!}{
	\begin{tabular}{c| c | c | c | c | c | c} %
		\toprule 
		\toprule 
				
		Dataset &  \begin{tabular}{@{}c@{}} $\#$ of  \\ labels \end{tabular}  & \begin{tabular}{@{}c@{}} Input  \\ size\end{tabular} & \begin{tabular}{@{}c@{}} $\#$ of  \\ samples \end{tabular} & \begin{tabular}{@{}c@{}} Model  \\ Architecture \end{tabular} & \begin{tabular}{@{}c@{}} Total  \\ Parameters \end{tabular} & \begin{tabular}{@{}c@{}} Total Model Accuracy \\ (Centralized data)  \end{tabular} \\ 
		\midrule
		ECG &  5 & 124 &  26,490 & \begin{tabular}{@{}c@{}} 4conv + 2dense \\ 1D CNN \end{tabular} & \begin{tabular}{@{}c@{}} 68,901 \end{tabular} & 97.78\%\\ 
		\hline
		\begin{tabular}{@{}c@{}} Speech Command \\ (SC) \end{tabular}  &  10 & 8,000 &  32,187 & \begin{tabular}{@{}c@{}} 4conv + 2dense \\ 1D CNN \end{tabular} & \begin{tabular}{@{}c@{}} 522,586 \end{tabular} & 85.29\% \\ 
		\bottomrule
	\end{tabular}
			}
	\label{tab:Setup} 
\end{table}


\section{Learning Performance Evaluation}\label{sec:learningEvaluation}

In practice, data is often distributed among clients in an imbalanced manner, e.g., some sensors are more active than others---with more data, and non-IID distributed, e.g., a single person's data can only be collected~\cite{li2019federated}. This situation corresponds to imbalanced data and non-IID data settings. Li et al.~\cite{li2019federated} studied the performance of FL under these settings. However, there is no work to analyze SplitNN in such a scenario yet.
Therefore, we are interested in conducting experiments based on the following research questions ({\bf RQ}).
\newline

\noindent{\bf RQ1: What factors/settings (e.g., number of clients, non-IID data, and imbalanced data) affect SplitNN learning performance?}
\newline

\noindent{\bf RQ2: Which setting will the SplitNN learning performance outperform FL?}


\subsection{IID and Balanced Dataset}\label{sec:idealDist}

Starting with ideal IID and balanced data distribution, we evaluate FL and SplitNN using both SC and ECG dataset\footnote{For all tests in this section if there is no explicit statement, the 4conv+2dense 1D CNN model architecture is used. The learning rate is set to be 0.001. The batch size = 32.}.

\begin{figure}[t]
	\centering
	\includegraphics[trim=0 0 0 0,clip,width=0.5\textwidth]{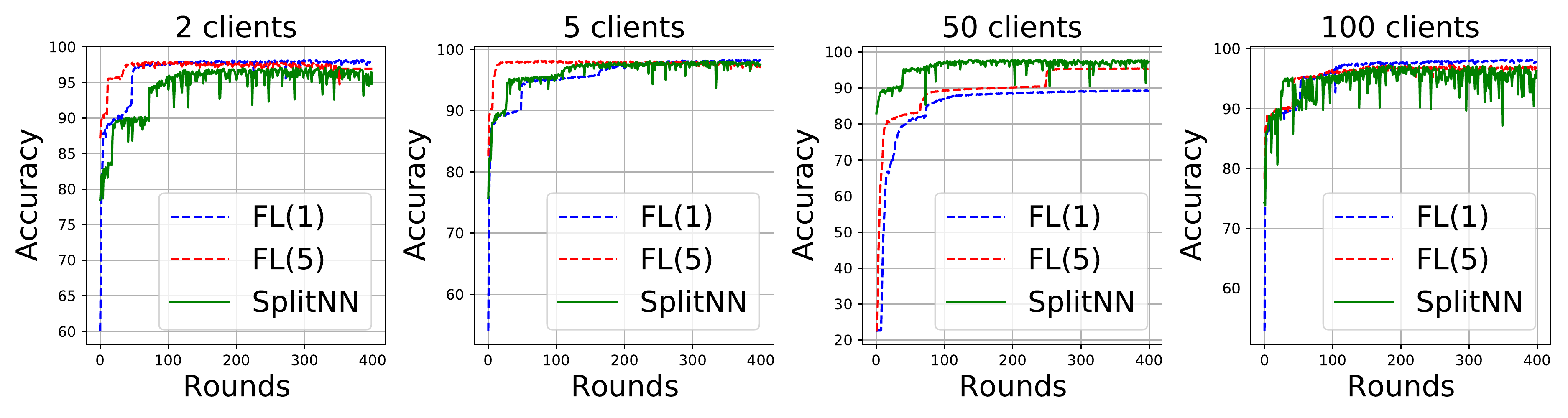}
	\caption{Testing accuracy of FL (1 and 5 local epochs for 1 round) and SplitNN over rounds for the ECG data, which is IID and distributed in a balanced manner.}
	\label{fig:IID_multiple_clients}
\end{figure}

\begin{figure}[t]
	\centering
	\includegraphics[trim=0 0 0 0,clip,width=0.5\textwidth]{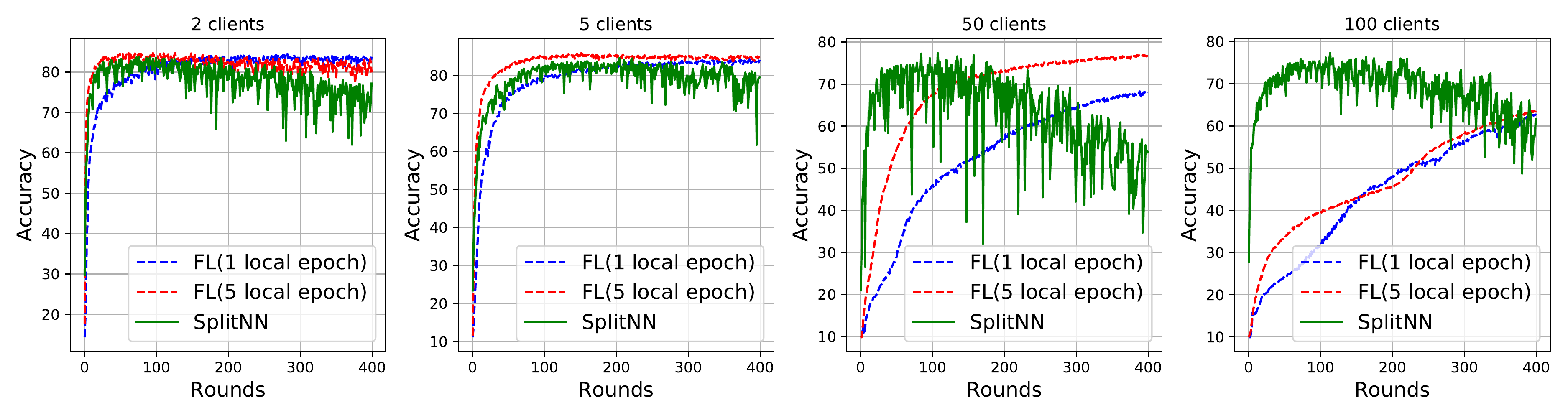}
	\caption{Testing accuracy of FL (1 and 5 local epochs for 1 round)  and SplitNN over rounds for the SC data, which is IID and distributed in a balanced manner.}
	\label{fig:IID_multiple_clients_sc}
\end{figure}

Fig.~\ref{fig:IID_multiple_clients} and~\ref{fig:IID_multiple_clients_sc} detail the testing accuracy over the number of rounds when FL and SplitNN are trained by a different number of clients---2, 5, 50, and 100 clients. Results indicate that SplitNN can always converge relatively faster than FL with one local epoch---notably for SplitNN, the local epoch is always 1. FL struggles with converging, especially when the number of clients becomes large. 


SplitNN testing accuracy starts dropping after it reaches an optimum point. Thus, an increasing number of rounds will not help improve accuracy. Stopping at the optimal point saves training time. In addition, SplitNN always exhibits an unstable learning curve with a high number of spikes. 

Furthermore, the model accuracy of SplitNN cannot reach the baseline accuracy of the centralized model---85.29\% for the SC and 97.78\% for the ECG, as detailed in Table~\ref{tab:Setup}. This limitation is clearly shown in Fig.~\ref{fig:IID_multiple_clients_sc}, when the number of clients is 50 or 100. 

These results indicate that the SplitNN model accuracy and convergence performance are not always the same as that of training a model through centralized data. Our findings are consistent with the previous conclusion in~\cite{gupta2018distributed}. However, we note that our findings are more generalized because we do not assume that the order of the data that arrived at multiple entities should be preserved, and the same initialization is used for assigning weights.


{\bf Remark1:} For {\bf RQ1}, SplitNN learning performance is affected by the number of clients. For {\bf RQ2}, SplitNN always outperforms the FL in terms of convergence speed in our experiments.

\subsection{Imbalanced Data Distribution}\label{sec:imbaDist}

\begin{figure}[t]
	\centering
	\includegraphics[trim=0 0 0 0,clip,width=0.5\textwidth]{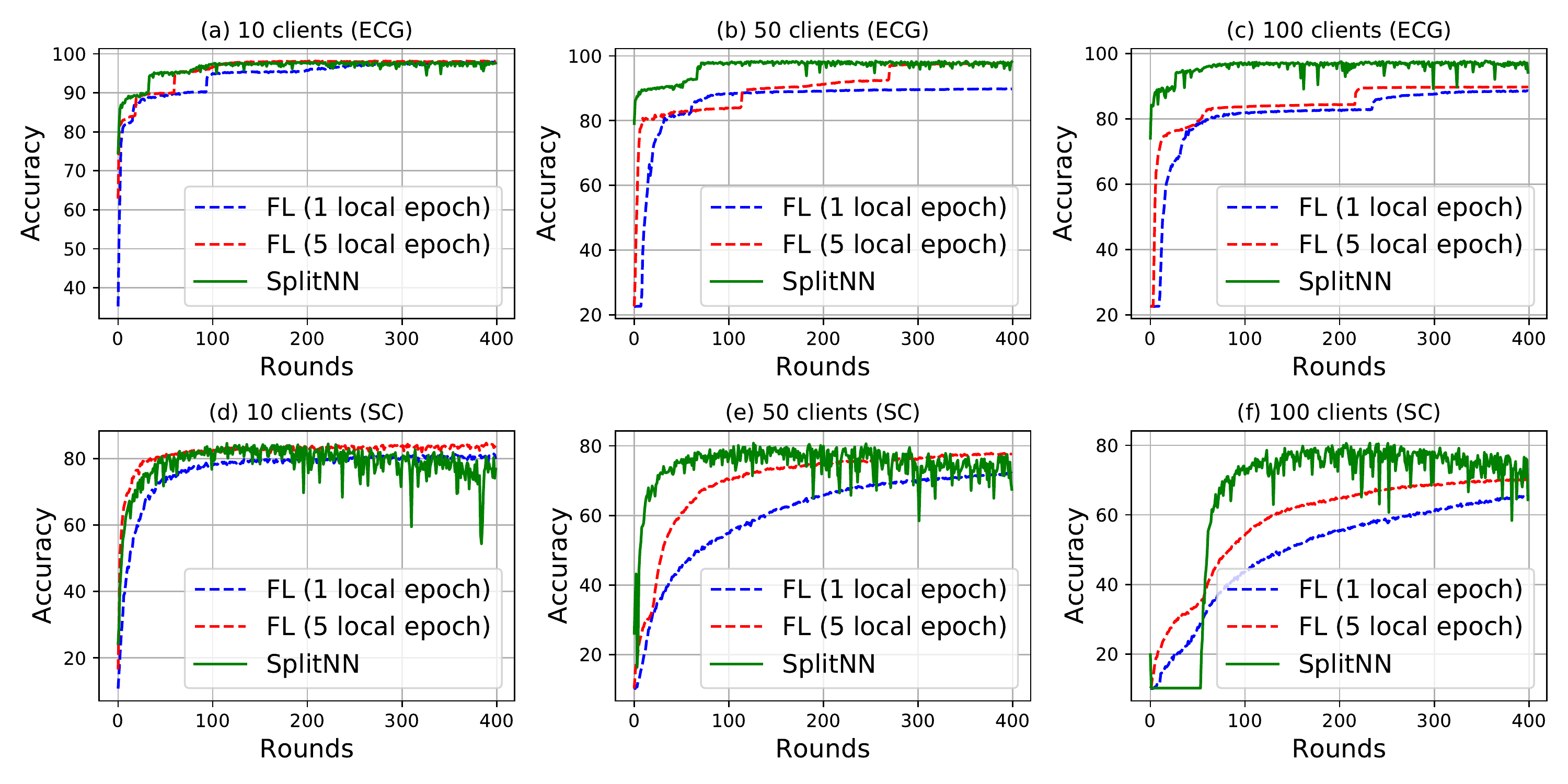}
	\caption{ Imbalanced data setting.}
	\label{fig:Imbalance}
\end{figure}

We assume the data are distributed among clients following the {\it normal distribution} to simulate the realistic imbalanced data distribution. Larger the sigma/variance, more imbalanced the data distributed. For example, when the number of clients is 10, and the total number of SC training dataset is 11360, the minimum number of training samples held by one client could be as few as 48 while the maximum number of training samples held by a client could be 3855---this is the setting for Fig.~\ref{fig:Imbalance} (d). We simulated clients up to 100. Given the same number of clients, {\it same data distribution} is applied to both FL and SplitNN.

According to Fig.~\ref{fig:Imbalance}, FL is hard to achieve the baseline accuracy of the centralized model, even when multiple local epochs per round is adopted for a large number of clients. For the SplitNN, its model accuracy deteriorates when the number of clients is large, e.g., Fig.~\ref{fig:Imbalance} (e) and (f). In addition, FL converges slower, especially when the number of clients goes up, e.g., 50 and 100 cases. Usage of more local epochs per round can expedite the convergence issue---but it cannot completely prevent---{\it given the similar communication overhead}. However, we note more local epochs proportionally prolongs training time on the client-side, although it can reduce the communication overhead. SplitNN is less sensitive to imbalance data distribution since it can always quickly converge. In Fig.~\ref{fig:Imbalance} (f), we can see that the training of SplitNN does not learn for the first 50 rounds/epochs. Once it starts learning, it indeed finds convergence quickly.

{\bf Remark2:} For {\bf RQ1}, the SplitNN learning performance is affected by both the number of clients and imbalanced data distribution. For {\bf RQ2}, SplitNN converges faster than FL in our experiments. 

\subsection{Non-IID Data Distribution}\label{sec:nonIIDDist}

\begin{figure}[t]
	\centering
	\includegraphics[trim=0 0 0 0,clip,width=0.5\textwidth]{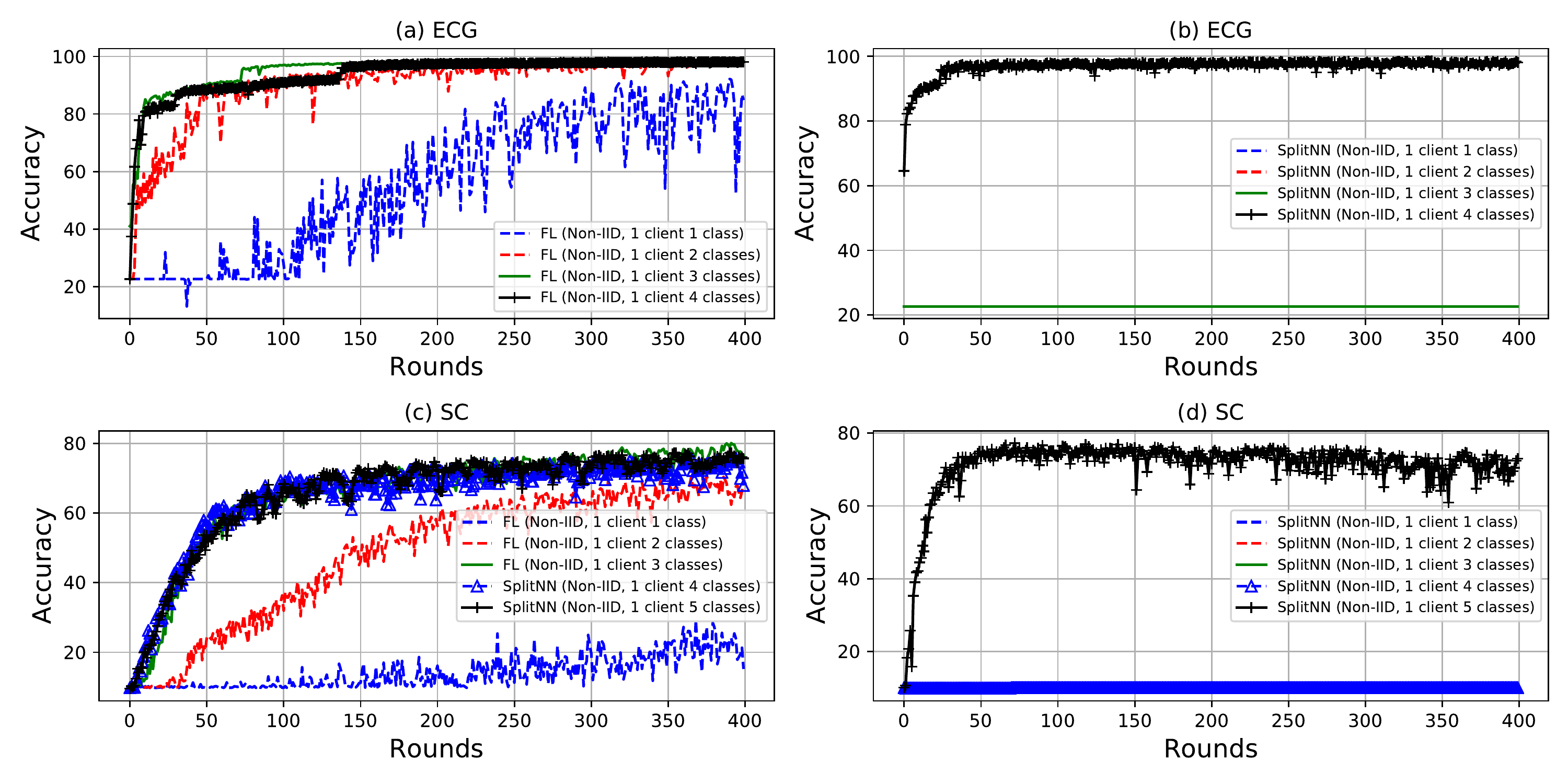}
	\caption{Non-IID dataset setting.} 
	\label{fig:nonIID}
\end{figure}


For the non-IID setting, the SC and ECG datasets are first sorted by class. Each client then receives data partition from only one single class, two classes, three classes, four classes, and five classes, respectively.

\noindent{\bf FL.} When each client has only one class data, FL convergence significantly fluctuates and is slow, as shown in Fig.~\ref{fig:nonIID} (a) and (c). It is clear from the figures that high skewness in the distributed data yields slower convergence and higher model accuracy drop. This finding is consistent with previous studies~\cite{zhao2018federated}. Nonetheless, FL can still converge in most cases, except 1 client with 1 class data.

\noindent{\bf SplitNN.} As shown in Fig~\ref{fig:nonIID} (b) and (d) that, unlike FL, which learns slowly under extreme non-IID distributed data, SplitNN does not learn at all. To be precise, as for the ECG with 5 total number of classes, the SplitNN model does not learn when one client holds 1 or 2 or 3 classes since the testing accuracy is always around 22.65\%, which is similar to guess (5 classes in total). As for the SC with 10 total number of classes, the SplitNN model does not learn on the condition when the client holds 1, 2, 3, or 4 classes. Therefore, in contrast to FL, SplitNN fails often to learn in non-IID settings. Note that in this experiments we have batch size = 32, and reducing the batch size (e.g., batch size = 4) may help SplitNN performance in case with one clients with 2 or 3 or 4 classes---but the trade-off is prolonged training time.

{\bf Remark3:} For {\bf RQ1}, SplitNN is very sensitive to non-IID data.
In fact, for both FL and SplitNN, some extend of knowledge forgetting while learning is evident when trained on the non-IID settings. Moreover, in our experiments, for {\bf RQ2}, FL outperforms SplitNN under non-IID data setting, especially extreme cases. The possible reason lies in the approach of model training. FL aggregates (averages) the local models trained on the local data present at each client. The aggregated model usually has more knowledge of the data classes even though there is one class per client than the cases where the model is sequentially learned over the clients without aggregation, such as in SplitNN.



\subsection{SplitNN enabled Ensemble Learning}

Here, we validate the SplitNN compatibility with ensemble learning. We use two different model architectures and simply train across two clients without loss of generality. Model M$_1$ is with 4 1D CNN layers and two dense layers, while M$_2$ is with 5 1D CNN layers and two dense layers. For both models, the client runs the first two CNN layers while the server runs the remaining layers.

\begin{figure}[t]
	\centering
	\includegraphics[trim=0 0 0 0,clip,width=0.48\textwidth]{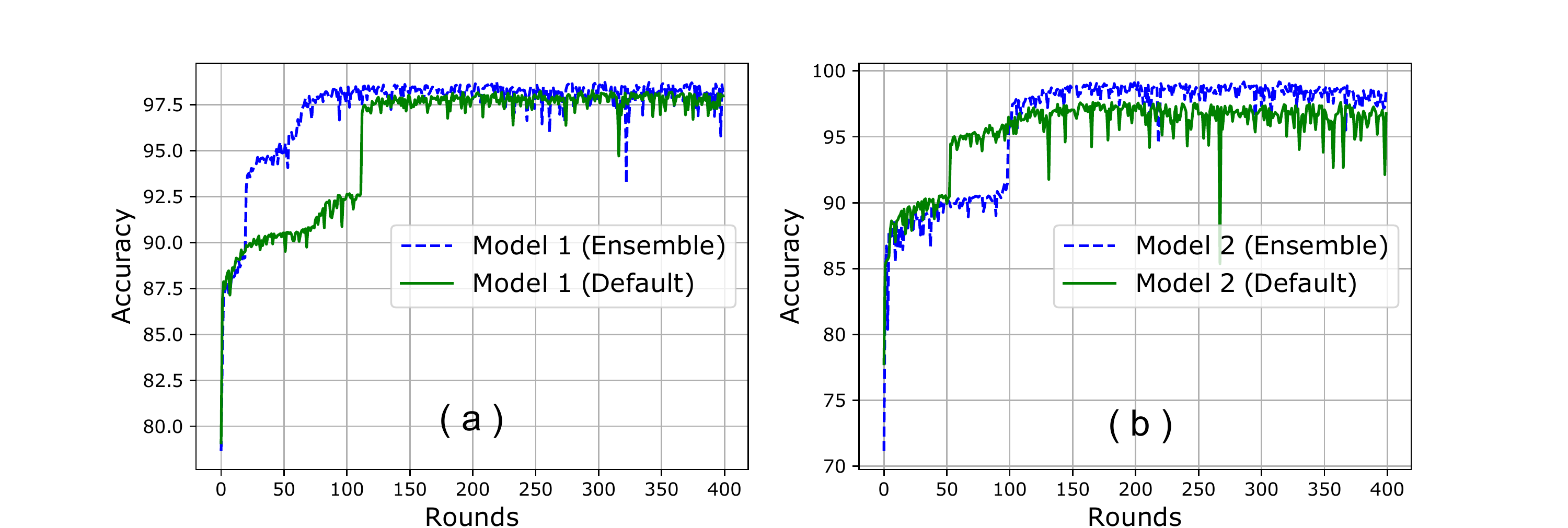}
	\caption{Ensemble learning that learns two models across two clients (ECG). (a) Model M$_1$, is with four 1D CNN layers and two dense layers. (b) Model M$_2$, is with five 1D CNN layers and two dense layers.}
	\label{fig:Ensemble_2client2Model}
\end{figure}

Both server and clients are simulated in the same desktop with one GTX1050 GPU and one i7-7700HQ CPU. Only the server uses the GPU, and the clients use CPU since the clients here are used to simulate IoT devices with low computational capability. Under this setting, the ensemble learning---following steps detailed in Section~\ref{sec:ensembleSetup}---takes 5178s in total to build two models. We further use SplitNN to train M$_1$ and M$_2$ across two clients individually---without ensemble. M$_1$ takes 3456s and M$_2$ takes 3625s. Therefore, the ensemble is faster to obtain two models---5178s versus 7081s (3456s + 3625s), which reduces the time overhead by 26.8\%. In terms of learning accuracy and convergence performance, there is no apparent difference when using ensemble learning to obtain two models and training them separately (marked as default), as depicted in Fig.~\ref{fig:Ensemble_2client2Model}. By taking advantage of the parallel computation capability of the server, multiple models can be obtained via SplitNN enabled ensemble learning. Notably, the server can further use distillation to obtain a single model from multiple models to achieve improved prediction accuracy and fewer variance~~\cite{hinton2015distilling}. 


\section{Implementation Overhead Evaluation on Raspberry Pi}\label{sec:implementationEvaluation}

\begin{figure}[h] %
	\centering
	\includegraphics[trim=0 0 0 0,clip,width=0.48\textwidth]{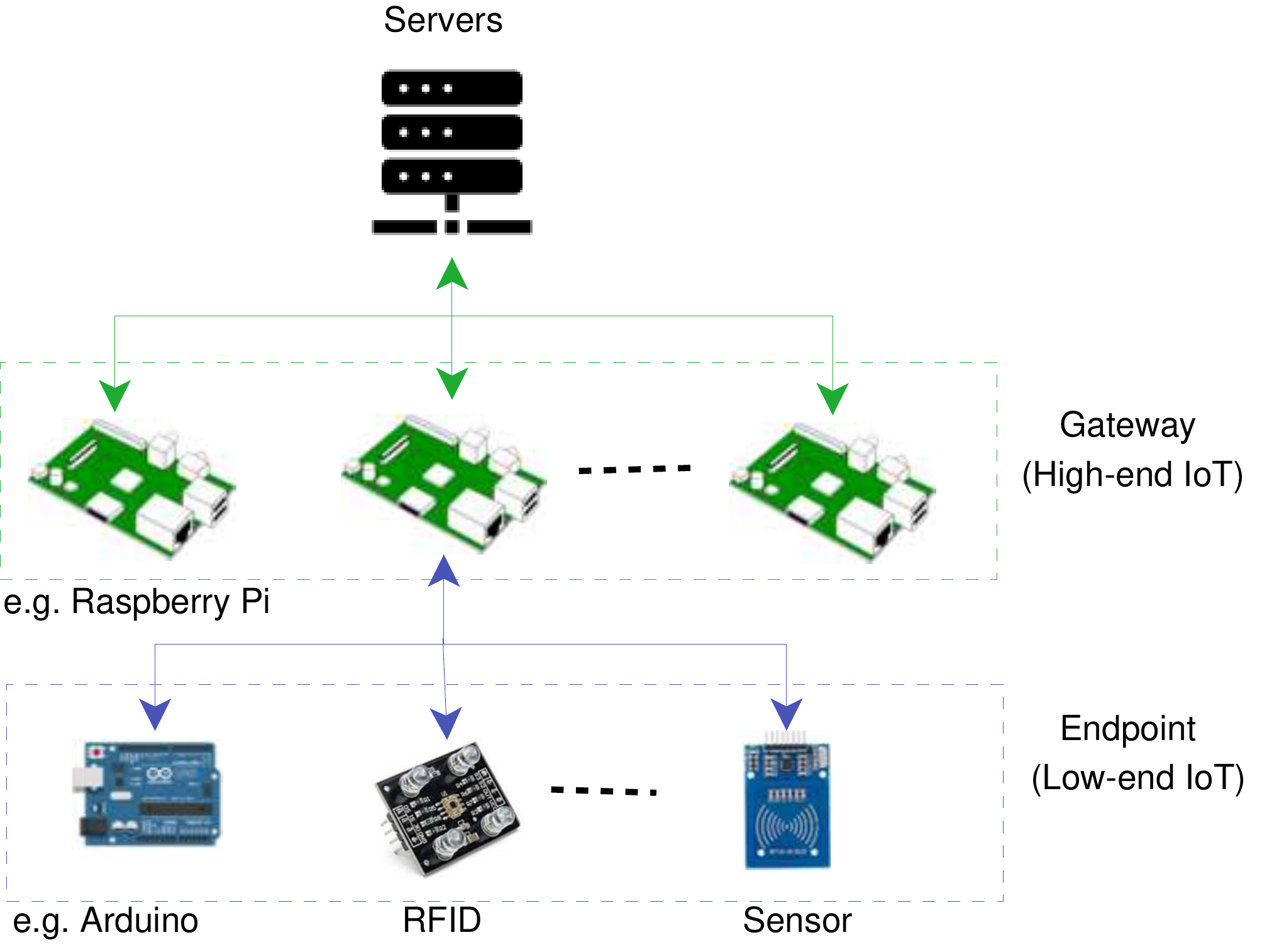}
	\caption{A typical IoT application setting. The IoT gateway (e.g., Raspberry Pi) aggregates data (e.g., from various IoT sensors) and interacts with the server to perform distributed learning.}
	\label{fig:IoTGateway}
\end{figure}

Using the ECG dataset, we evaluate time, power, communication, and memory overhead when running FL and SplitNN on real IoT devices, Raspberry Pi, to provide a benchmark under real-world IoT settings. In particular, we simulate one typical IoT application scenario, as illustrated in Fig.~\ref{fig:IoTGateway}, similar to~\cite{nguyen2019diot}, which can be a smart home setting.
According to~\cite{musaddiq2018survey}, the IoT device can be generally categorized to \emph{high-end} IoT device and \emph{low-end} IoT device. The low-end IoT devices are temperature, motion sensors, and RFID cards, which are usually strictly resource-constraint. They may not even support an OS such as Linux to run a machine learning algorithm. High-end IoT devices are simple devices like Raspberry Pi. Hence, in this simulated IoT application scenario, Pi serves as a gateway, which aggregates data from low-end IoT devices, e.g., sensors, and interacts with the server to perform distributed learning tasks. 

We considered the following test settings\footnote{We set \textit{one local epoch per round for FL} in all experiments. We compare implementation overhead by presetting a fixed number of \textit{100 rounds} for both FL and SplitNN. In other words, 100 epochs for both of them. We always use the learning rate of 0.001.}:

\begin{enumerate}
    \item Ensemble learning to train model M$_1$ and M$_2$ concurrently across 2 clients, as well as training M$_1$ and M$_1$ individually across two clients via SplitNN (Section~\ref{sec:resultEnsemble});
    \item Evaluating FL and SplitNN across a range of clients from two to five with the same model architecture (Section~\ref{sec:resultDifClient});
    \item Evaluating FL and SplitNN across five clients with different model architectures. For the SplitNN, two split layers run on clients regardless of model architectures (Section~\ref{sec:resultDifSplit});
    \item Evaluating SplitNN when a different number of layers is split and running on the client, given the same model architecture. Specifically, one, two, three layers are split and running on the client (Section~\ref{sec:resultDifModels}).
\end{enumerate}

In the experiments, we use one Raspberry Pi device to act as the IoT gateway.


\subsection{Experimental Setup}

\begin{figure}[!ht]
	\centering
	\includegraphics[trim=0 0 0 0,clip,width=0.5\textwidth]{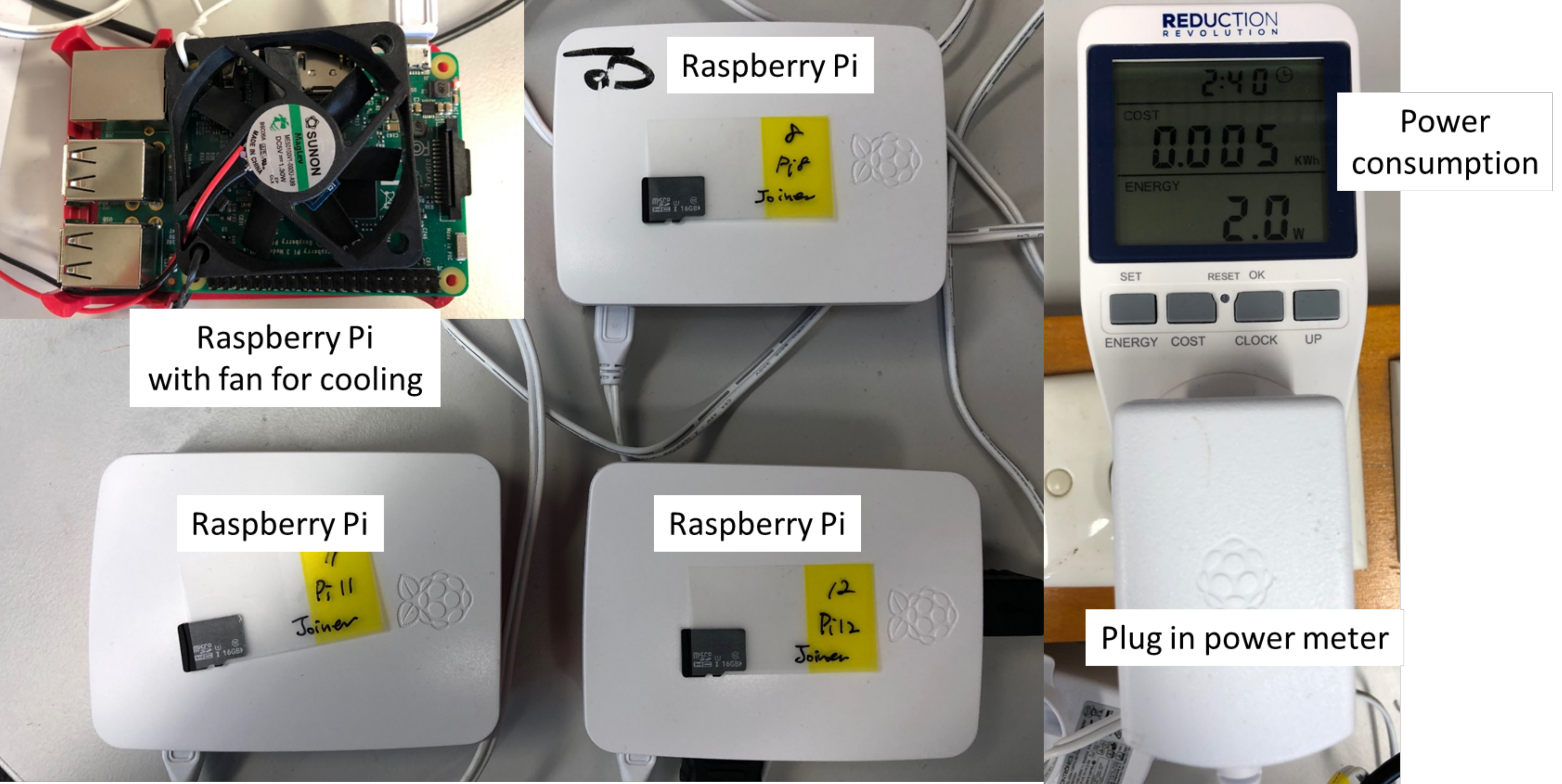}
	\caption{Four Raspberry Pi devices and a power meter are shown. 
	See the demo video for more details, \url{https://www.youtube.com/watch?v=x5mD1_EA2ps}.}
	\label{fig:setup}
\end{figure}

We use the Raspberry Pi 3 model BV1.2 (Fig.~\ref{fig:setup}) with the following settings: PyTorch version 1.0.0, OS Raspbian GNU/Linux 10 (buster), and Python version 3.7.3. We note that CUDA is not available for the model. The server (laptop) has the following settings: CPU i7-7700HQ, GPU GTX 1050, Pytorch version 1.0.0, OS windows 10, Python version 3.6.8 using Anaconda, and the CUDA version 10.1.

\subsection{Measurement Methods of Performance Metrics}

\paragraph{Training Time} 

We use Python's \textsf{time} library to measure the training time containing the communication time between client and server. We set the time as $T_{\rm start}$ when the model starts training. Once the training is finished, we set the time as $T_{\rm end}$. As a result, training time is $T_{\rm end} - T_{\rm start}$. 

\paragraph{Memory Usage}
We use Linux \texttt{free -h} command for measuring the memory usage. This command provides the memory information of \texttt{total}, \texttt{used}, \texttt{free}, \texttt{cached} and \texttt{available}. 
The total memory of the Raspberry Pi device used in this experiment is 926~MB. The focus here is to record and report the \texttt{used} memory during training.

\paragraph{Power Consumption}

We use a plug-in powermeter, as shown in Fig.~\ref{fig:setup}, to measure the power consumption. We measured the power consumption in the kilowatt-hour (kWh) unit.   

\paragraph{Temperature}

We use Python's \textsf{CPUTemperature}() function from the \texttt{CPUTemperature} library to monitor the temperature of the Raspberry Pi CPU.

\paragraph{Communication Overhead}
We measure the transmitted data size from each client to the server and vice versa. We use the \texttt{pickle} library to monitor the size of the transmitted data. We use the router DGN2200 v4 (N300 Wireless ADSL2+ Modem Router) for wireless communication between Raspberry Pi and the server.

\subsection{Implementation Considerations}\label{sec:engineer}

\subsubsection{Low Performance} Although Raspberry Pi is regarded as a high-end IoT device~\cite{musaddiq2018survey}, its computational resources are still quite limited compared with traditional computing devices such as servers and PCs. We first tried to run MobileNet~\cite{howard2017mobilenets}. When we trained CIFAR10 dataset with MobileNetv1\footnote{Source code is adopted from \url{https://github.com/Tshzzz/cifar10.classifer/blob/master/models/mobilenet.py}.} (20 conv2D layers with 3,228,170 model parameters in total), it took 8 hours 41 minutes for FL per round with 1 local epoch. For SplitNN, it took about 2.5 hours for one epoch across five Raspberry Pi devices\footnote{Since the training sample for CIFAR10 is 50,000, we use 5 clients. Therefore, each client holds 10,000 images for both FL and SplitNN.}, when only the first two layers are running on the Raspberry Pi device.   
In addition, we tried to run ResNet20\footnote{Source code is adopted from \url{https://github.com/akamaster/pytorch_resnet_cifar10/blob/master/resnet.py}.} (with 20 conv2D layers and 269,722 parameters). When we ran SplitNN across 5 Raspberry Pi devices, it took about 1 hour to finish one round when the first two layers are only running on the Pi devices. When we also ran FL across 5 Raspberry Pi devices, it took 37 minutes for one round with one local epoch.



Based on those results, it seems challenging to run either ResNet20 or MobleNet V1 on Raspberry Pi devices. Because those models have several million parameters, they might be computationally heavy for simple IoT devices even though these 2D CNN models are known as light models for the GPU platform\footnote{This does not mean {\it inference task} cannot be properly performed by Rasberry Pi given the model already trained and optimized.}. Therefore, for full end-to-end tests, we opt for evaluating a relatively simple model, such as 1D CNN, on sequential time series data. This evaluation would be valuable for IoT settings because sequential time series data are pervasive data sources generated and collected by various IoT devices, e.g., sensors.


\subsubsection{Install Pytorch on Raspberry Pi}
We have made a unified manual guide of installing Pytorch v1.0.0 on Raspberry Pi\footnote{This manual is available via \url{https://github.com/Minki-Kim95/RaspberryPi}.}. We believe that this manual will help developers because we explain how to address the errors during installation, which are hard to resolve, and there are no solutions online.
\subsubsection{Temperature of Raspberry Pi}
When training on the Raspberry Pi device, especially FL that runs the entire model on the device, the temperature goes high---usually more than 80$\celsius$. Therefore, it is necessary to cool down the device. A cooling fan can be attached to the Raspberry Pi (Fig.~\ref{fig:setup}). This practice can efficiently cool it down from 83$\celsius$ to 54$\celsius$.

\subsection{Ensemble Learning}\label{sec:resultEnsemble}
\begin{figure*}[h]
	\centering
	\includegraphics[trim=0 0 0 0,clip,width=1.0\textwidth]{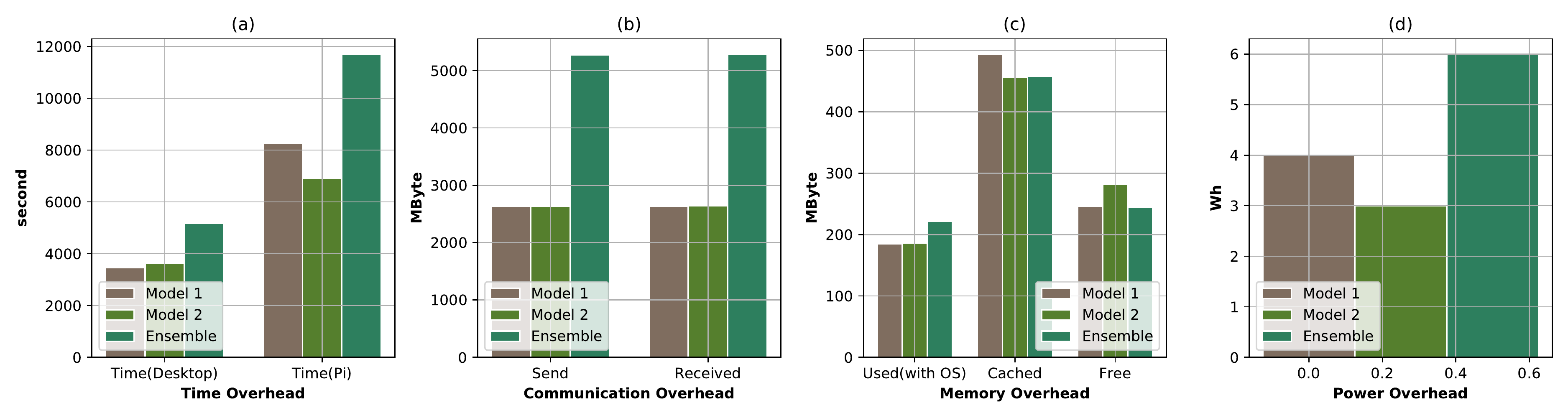}
	\caption{Ensemble learning vs learning model individually (100 rounds are commonly used for both approaches.).}
	\label{fig:Ensemble}
\end{figure*}

If we train M$_1$ with $T_1$ and $T_2$, respectively, the SplitNN enabled ensemble training should roughly take \textsf{max}($T_1$, $T_2$) on the condition that the server has unlimited computational power because the server can train the remaining layers of M$_1$  and M$_1$ in parallel. As we treat the laptop equipped with only a low-end GPU as a server, this ideal case is not met. Because the laptop is unable to ideally parallel train the remaining layers of M$_1$  and M$_2$. But, as shown in Fig.~\ref{fig:Ensemble} (a), the time for ensemble training, $T_{\rm ensem}$ is indeed always smaller than sequentially train each model, $T_1 + T_2$. Specifically, $T_{\rm ensem}$ takes 11704 seconds to concurrently obtain both M$_1$ and M$_2$. Training M$_1$ and M$_2$ individually costs 8267 seconds and 6908 seconds, respectively. Therefore, ensemble learning reduces the time overhead by 22.87\%. For the communication overhead, ensemble learning is the same as individually training each model (Fig.~\ref{fig:Ensemble} (b)). For the memory usage, we rely on the \texttt{used} memory for comparison\footnote{This includes 119MB memory used by the OS by default as it tends to be fair to include the memory occupied by the OS.}. Used memories of individually training M$_1$ and M$_2$ are 185MB and 186MB. In contrast, the used memory of ensemble training is 222MB. 
For power consumption, sequentially training M$_1$ and M$_2$ consumes 4Wh (watt per hour) and 3Wh, while ensemble training consumes 6Wh. Therefore, ensemble learning can reduce power consumption, as well. 

In summary, ensemble learning does not reduce communication overhead. However, it has the potential to reduce training time, used memory, and power consumption. 

\subsection{Effects of Number of Clients}\label{sec:resultDifClient}

Here, we evaluate both FL and SplitNN when the number of Raspberry Pi devices (clients) varies from two to five. 
This experiment focuses on the overhead brought to individual Raspberry Pi rather than the server side. One can easily extend the number of clients beyond five by adopting the released artifact (source code, user guide and demo) of our experiment.

The model architecture has four 1D CNN layers and two dense layers. For SplitNN, the first two 1D CNN layers run on Raspberry Pi devices.

\begin{figure*}[h]
	\centering
	\includegraphics[trim=0 0 0 0,clip,width=0.95\textwidth]{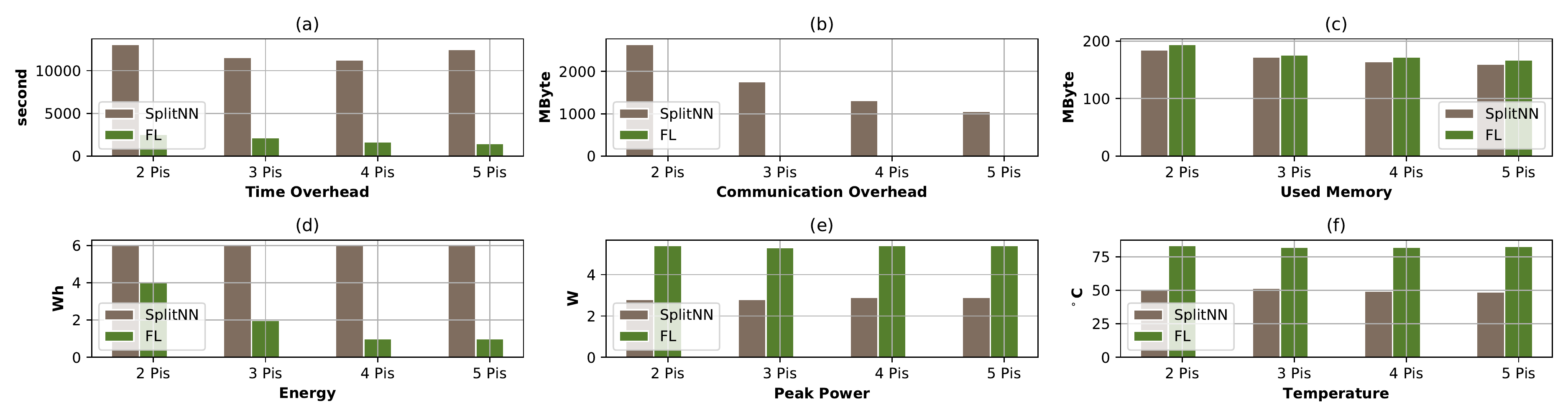}
	\caption{FL and SplitNN evaluation when the number of Raspberry Pi devices (clients) varies from two to five. All tests were performed with the 10 Gbit/s dedicated LAN.}
	\label{fig:FLVsSplitNN_Pis_ecg}
\end{figure*}

We report the performance overhead for a single Raspberry Pi device because we are interested in the client's overhead. The performance results are presented in Fig.~\ref{fig:FLVsSplitNN_Pis_ecg}.  
As for the time overhead in Fig.~\ref{fig:FLVsSplitNN_Pis_ecg} (a), FL reduces as the number of devices increases. This is due to the decrease in the local data size. SplitNN slightly increases since each device runs the training sequentially. Overall, SplitNN usually takes several times longer than the FL, given the same number of rounds.

The communication overhead is presented in Fig.~\ref{fig:FLVsSplitNN_Pis_ecg} (b)\footnote{We note that the communication overhead of FL is not displayed because it is orders of magnitudes smaller than that of SplitNN, e.g., megabytes vs. gigabytes.}. FL stays relatively constantly around 28,552,161 bytes because the model parameters determine the FL's communication overhead rather than the local data size. SplitNN communication overhead decreases as it is highly related to the local data size. This corroborates with the statistical analysis result in a recent work~\cite{singh2019detailed}, where the communication overhead of SplitNN is shown significantly higher than that of FL per round for low model complexity and fewer clients.


For the used memory, as shown in Fig.~\ref{fig:FLVsSplitNN_Pis_ecg} (c), the FL is always higher than that of SplitNN, because the FL needs to train the entire model in the Raspberry Pi, while the SplitNN only needs to train a small part---few split layers. This also leads to the high power peak, as shown in Fig.~\ref{fig:FLVsSplitNN_Pis_ecg} (e), and high temperature during training, as shown in Fig.~\ref{fig:FLVsSplitNN_Pis_ecg} (f), in the FL case. Without cooling, the Raspberry Pi device's temperature can be up to $83\celsius$ during the FL learning. 

However, although FL has a high power peak, the energy is lesser than that of SplitNN for the same number of rounds, as shown in Fig.~\ref{fig:FLVsSplitNN_Pis_ecg} (d). This is because, in FL, each client trains local model {\it in parallel}, and consequently, the total time (accumulated computation and communication) for running a given number of rounds is less than that of SplitNN.




\begin{figure*}[h]
	\centering
	\includegraphics[trim=0 0 0 0,clip,width=1.0\textwidth]{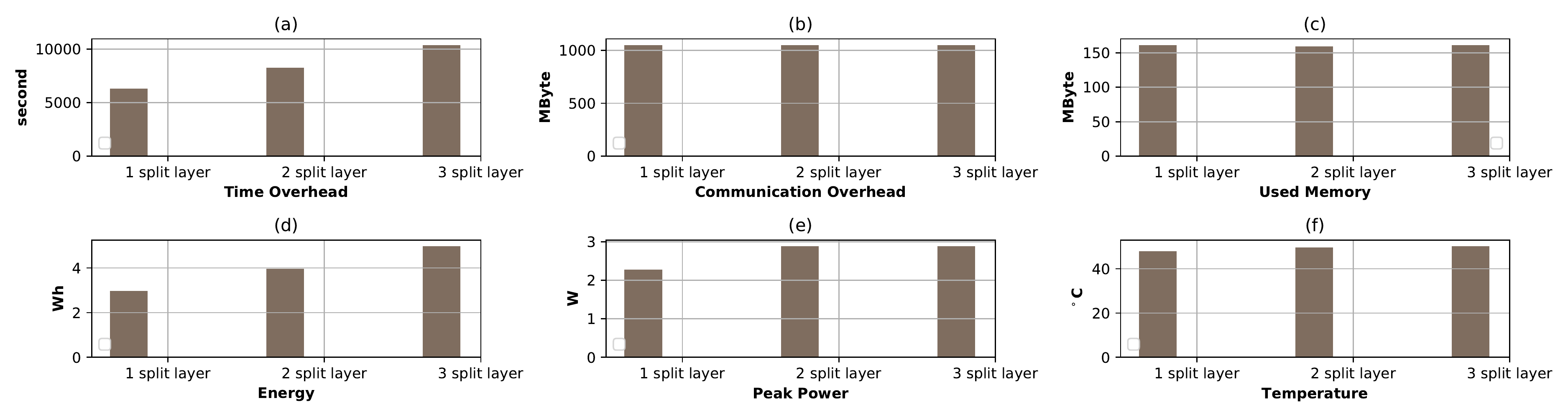}
	\caption{SplitNN performance when a different number of split layers---1 to 3 convolutional layers---run on Raspberry Pi devices. All tests run {\it across 5 Raspberry Pi devices} and in a lab environment equipped with the 100 Gbit/s dedicated LAN. The model is with 4 convolutional layers and 2 dense layers.}
	\label{fig:PI_FLVsSplitNN_differentLayers_ecg}
\end{figure*}

\subsection{Effects of Number of Split Layers in SplitNN}\label{sec:resultDifSplit}

Experiments are carried on five Raspberry Pi devices to observe the effect of the number of split layers for SplitNN. We note that this experiment is only applicable to SplitNN because FL has to train the entire model on each client.

As shown in Fig.~\ref{fig:PI_FLVsSplitNN_differentLayers_ecg} (b), the communication overhead remains the same regardless of the number of layers at the client-side because the communication overhead in SplitNN depends on the number of parameters in the cut layer rather than the number of split layers. As for the memory usage in Fig.~\ref{fig:PI_FLVsSplitNN_differentLayers_ecg} (c), we observe only a slight increase with the number of split layers. Most noticeably, time overhead (depicted in  Fig.~\ref{fig:PI_FLVsSplitNN_differentLayers_ecg} (a)) and energy overhead (depicted in Fig.~\ref{fig:PI_FLVsSplitNN_differentLayers_ecg} (d)) increase with the number of split layers because the number of parameters to be trained on each client increases. Therefore, in practice, from the overhead reduction perspective, it is preferred to run a few layers only at the client-side for SplitNN. 


\subsection{Effects of Different Models}\label{sec:resultDifModels}

To observe the effects of different models for SplitNN and FL, we perform experiments on five Raspberry Pi devices. The models have a varying number of convolutional layers ranging from four to eight. For SplitNN, we use a fixed number of split layers running on each client regardless of the number of layers in the entire model. 

\begin{figure*}[h]
	\centering
	\includegraphics[trim=0 0 0 0,clip,width=1.0\textwidth]{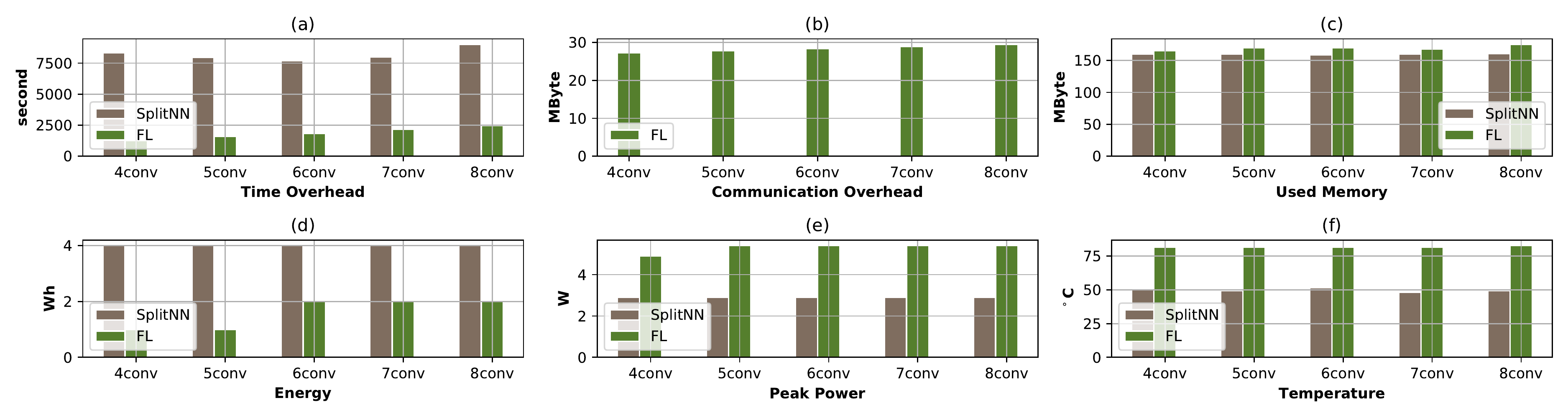}
	\caption{Overhead performance of FL and SplitNN when the number of layers of the model varies from 4 conv to 8 conv. All tests were performed with the 100 Gbit/s dedicated LAN. For SplitNN, the first two convolutional layers run at the client-side. For communication overhead, we only show the results for FL because SplitNN's communication overhead (1054Mbytes) is not changed as well as significantly greater than FL's communication overhead, as shown in Fig.~\ref{fig:FLVsSplitNN_Pis_ecg}.}
	\label{fig:PI_FLVsSplitNN_differentModel_ecg}
\end{figure*}

According to results depicted in Fig.~\ref{fig:PI_FLVsSplitNN_differentModel_ecg}, the overhead, including time, communication, memory used, and energy consumed by Raspberry Pi devices (linearly) increases with the model complexity (defined by the number of layers in the model) for FL. In contrast, the overhead remains more or less constant for SplitNN because the number of layers running on each client is fixed. Based on these findings, SplitNN becomes more advantageous when we consider a complicated model.


\section{Discussion and Future Work}\label{sec:discussion}

\subsection{Summary of the Evaluation Results}
Firstly, we evaluated the learning performance of both FL and SplitNN in terms of model accuracy and convergence speed. For {\bf RQ1} and {\bf RQ2}, our experimental results provide following insights:
\begin{enumerate}
    \item SplitNN learning performance is indeed sensitive to various settings, including i) the number of clients, ii) non-IID, and imbalanced data distribution.
    \item In comparison with FL, SplitNN always converges much faster under imbalanced data distribution. However, SplitNN is more sensitive to non-IID data, where it even does not learn at all under extreme non-IID cases, e.g., one client with 2 classes or one client with 3 classes.
\end{enumerate}


Next, we fully implemented both FL and SplitNN on Raspberry Pi devices using 1D CNN models with a small number of parameters to evaluate the performance of both models in the case of IoT devices with low computational capability. Our experimental setup simulates a practical scenario for edge distributed learning. Our experimental results on those devices suggest the following findings: 
\begin{enumerate}
    \item If we consider the communication cost as the most critical metric in IoT applications, FL is preferred over SplitNN because FL has relatively lower communication overhead. Interestingly, we found that reducing communication overhead can also benefit training time and energy consumption.
    \item For applications where the communication is not a significant concern (e.g., Ethernet or 5G are available), SplitNN is recommended to ensure better model accuracy and guarantee (fast) convergence except the case of (extreme) non-IID data distribution.
\end{enumerate}

\subsection{Optimization of the Implementation}

This work follows a typical setting of FL and SplitNN, and optimization is out of scope, especially when mounting on the IoT devices. We empirically found that {\it training} 2D CNN models, such as ResNet and MobileNet, on Raspberry Pi devices is computationally infeasible. For building a more memory and computation efficient model, one possible optimization is to use XNOR-NET~\cite{rastegari2016xnor,mcdonnell2018training}. We can also consider training convolutional neural networks using addition operations only without multiplication operations based on the previous study results~\cite{chen2019addernet} since multiplication operations are significantly computationally heavier than addition operations in CNN models. Besides, to reduce the training overhead due to sequential training in SplitNN, a parallel machine learning model update paradigm of FL is applicable in its client-side section~\cite{splitfed}.  

\subsection{Splitting Sequential Models}

To process sequential time-series data for experiments, we used a 1D CNN model because 1D CNN models can be split vertically, which is easy to apply SplitNN. Actually, before choosing the 1D CNN to deal with sequential data, we tried to apply SplitNN on other sequential models such as LSTM and RNN that are popularly used state-of-the-art machine learning models to deal with sequential data. However, we found that such sequential models are hard to be applied to SplitNN\footnote{FL applies to sequential models.}. Therefore, the applicability of SplitNN for sequential models leaves future work.

\section{Conclusion}
\label{sec:Conclusion}

This work is the first to empirically evaluate SplitNN and compare it with FL in real-world IoT settings. We comprehensively evaluated the learning performance in terms of model accuracy and convergence speed of FL and SplitNN. For our experiments, we mainly considered imbalanced and non-IID distributions, which would be more suitable for IoT scenarios. Similar to FL, SplitNN is also inevitably influenced by data distribution. In general, SplitNN performs better than FL in the case of imbalanced data distributions but can rather worsen than FL in the case of extreme non-IID data distributions.
Beyond empirical learning performance evaluation and comparison, we extensively evaluated the practicality of mounting FL and SplitNN on Raspberry Pi devices to simulate real-world IoT scenarios. We mainly dealt with pervasive sequential time-series data and provided useful comprehensive results---various implementation overhead---to the community. Overall, for the IoT scenario, the FL would be a more practical recommendation because it requires less overall communication, time, and power consumption overhead when a simple 1D CNN model is used. However, we also found that for both FL and SplitNN, the use of more complicated models would still be infeasible to mount {\it training} on low-capacity IoT devices such as Raspberry Pi.

\section*{Acknowledgment}
The work has been supported by the Cyber Security Research Centre Limited whose activities are partially funded by the Australian Government’s Cooperative Research Centres Programme. This work was also supported in part by the ITRC support program (IITP-2019-2015-0-00403). The authors would like to thank all the anonymous reviewers for their valuable feedback.
%

\end{document}